\pdfoutput=1

\documentclass{article}
\usepackage{amsmath}
\usepackage{amssymb}
\usepackage{graphicx}

\newcommand\Ro[1][\relax]{\ifx\relax#1 \ensuremath{\mathcal{R}_0}
  \else \ensuremath{\mathcal{R}_{0,#1}} \fi}
\newcommand{\order}{\mathcal{O}}

\title{Epidemics with general generation interval distributions}
\author{Joel C. Miller, Bahman Davoudi, Rafael Meza, Anja Slim,\\ Babak Pourbohloul}

\newcommand{\f}{F}
\newcommand{\g}{G}
\newcommand{\h}{H}
\newcommand{\F}{\mathcal{F}}
\newcommand{\G}{\mathcal{G}}
\renewcommand{\H}{\mathcal{H}}

\begin{document}
\maketitle
\begin{abstract}
  We study the spread of susceptible-infected-recovered (SIR)
  infectious diseases where an individual's infectiousness and
  probability of recovery depend on his/her ``age'' of infection.  We
  focus first on early outbreak stages when stochastic effects
  dominate and show that epidemics tend to happen faster than
  deterministic calculations predict.  If an outbreak is sufficiently
  large, stochastic effects are negligible and we modify the standard
  ordinary differential equation (ODE) model to accommodate
  age-of-infection effects.  We avoid the use of partial differential
  equations which typically appear in related models.  We introduce a
  ``memoryless'' ODE system which approximates the true solutions.
  Finally, we analyze the transition from the stochastic to the
  deterministic phase.
\end{abstract}

\section{Introduction}
\label{sec:introduction}
Infectious diseases continue to impact public health.  The previous
emergence of SARS, the ongoing emergence of H1N1 swine influenza, and
the simmering threat of H5N1 avian influenza or other diseases call
attention to the need to prepare for a quickly-spreading pandemic.
Such a pandemic could have typical infectious period measured in days
or weeks, spread worldwide, and grow quickly.  In the face of such an
emerging disease, there is little time to develop and implement
interventions.

The ability to predict the timing and maximum patient load imposed by
an epidemic is essential to intervention design.  Overestimating the
preparation time available or underestimating the peak may result in
well-designed measures which are implemented too late or are too
small.

The ability of an infectious disease to spread depends strongly on the
number of susceptible individuals $S$, and the total population size
$N$.  We will find that the details of the spread are more sensative
to changes in $N/S$ than changes in $S/N$, and so we will couch most
of our discussion in terms of changes in $N/S$.

\begin{figure}
\includegraphics[width=0.32\textwidth]{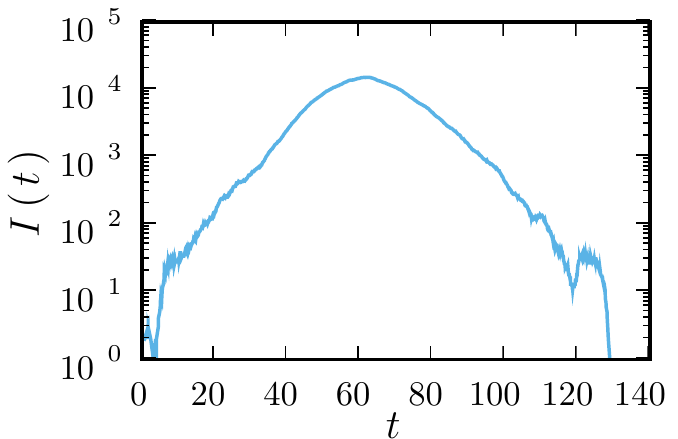} \includegraphics[width=0.32\textwidth]{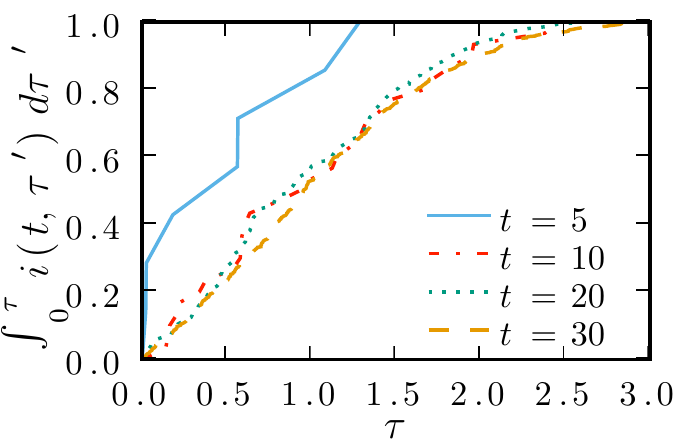} \includegraphics[width=0.32\textwidth]{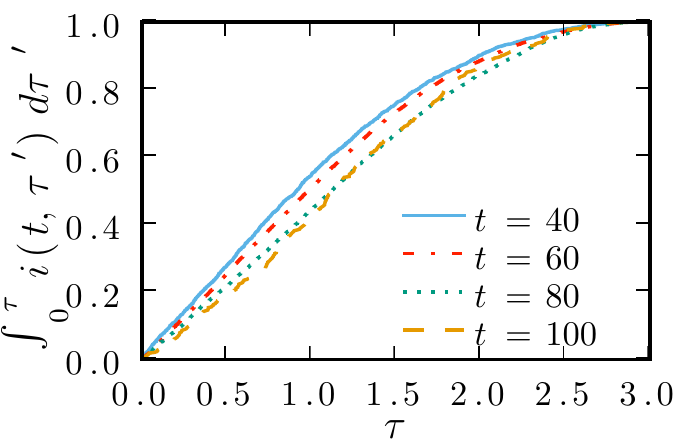}
\caption{The course of an epidemic with vertical logscale (left).  The
  cumulative age-of-infection distribution $\int_0^\tau i(t,\tau')\, \mathrm{d}\tau'$ at
  different times (center and right).}
\label{fig:sample}
\end{figure}

Figure~\ref{fig:sample} shows the course of an epidemic of an infectious
disease whose characteristics are discussed later
(\S{}\ref{sec:pc} with $c=0.9$).  At very early times the disease
spreads as a branching process and stochastic effects are important.
As the outbreak grows, the spread continues as a branching process,
but the stochastic effects lose importance.  However, the timing of
the epidemic always feels the initial stochastic impact.

We also consider the infection-age distribution $i(t,\tau)$, the
number of people infected at time $t$ who have been infected for
$\tau$ units of time.  We plot the cumulative distribution in
figure~\ref{fig:sample} at small $t$ (center) and larger $t$ (right).
At small $t$ the distributions are noisy, and converge to a
steady-state distribution as $t$ increases.  As the spread continues,
$N/S$ begins to change perceptibly and the steady-state adjusts
adiabatically if $N/S$ changes slowly enough.  If $N/S$ does not
change slowly, the system cannot adjust to the changing equilibrium.
During the growing phase of the epidemic, the infected individuals are
weighted towards more recent infections, while during the declining
phase the infected individuals have disproportionately older
infections.

We focus on several stages in this paper: the early stochastic phase,
the later deterministic phase, and the transition phase between these
two.  If $S$ is initially small, then $N/S$ can change significantly during the
stochastic phase.  We do not address this case.

Typically disease outbreaks are either subcritical (meaning $\Ro<1$)
for which epidemics are impossible because an average infected person
infects fewer than one individual, or supercritical (meaning $\Ro>1$)
for which epidemics are possible.  We consider only supercritical
outbreaks.  Early in an outbreak's spread, growth is dominated by
stochastic effects, and it may die out stochastically.  If it
persists, it may grow faster or slower than ``average''.  As long as $N/S$ does not change significantly,
the spread can be modeled using Crump-Mode-Jagers (CMJ)
processes~\cite{CM1,CM2,jagers1,jagers2}.  A
subcritical CMJ process dies out, while a supercritical CMJ process
either dies out or converges to $Ce^{\phi t}$ where $C$ is a random
value and $\phi$ depends on the process.

If a supercritical outbreak becomes sufficiently large the spread is
effectively deterministic.  The usual equations for this phase
are the susceptible-infected-recovered (SIR) equations
\begin{align}
\dot{S} &= -\beta IS/N \label{eqn:Sdot}\\
\dot{I} &= \beta IS/N -\gamma I\\
\dot{R} &= \gamma I \label{eqn:Rdot}
\end{align}
These equations assume that infected people cause infections at rate
$\beta$ and recover at rate $\gamma$, giving an exponentially
distributed infection duration.  The process is ``memoryless''.  In
contrast, for real diseases the ``age'' of an individual's infection
affects his/her infectiousness and probability of recovering.

Ignoring ``age-of-infection'' effects loses important details.
During the growth of an epidemic the infections are biased towards
young infection ages.  If young infections are more (or less)
infectious, the SIR equations under- (or over-) estimate the growth
rate.  Similar observations hold during decay.

Several approaches have been developed to study age-of-infection
models.  Some explicitly track the history of the
epidemic~\cite{breban:aoi,hethcote:aoi,brauer:aoi,brauer:compartmental,li:aoi,castillochavez:aoi,thieme:aoi}.
Others attempt to maintain the memoryless feature of
equations~\eqref{eqn:Sdot}--\eqref{eqn:Rdot} by introducing a chain of
infected compartments $I_1, \ldots, I_n$ in order to approximate the
infectious period
distribution~\cite{anderson:gammadist,wearing,ma,babak:gammadist,lloyd:destabilization,lloyd}.
These chains of compartments usually do not have biological meaning,
but instead are a simplifying ``trick''.  Typically these assume
constant $\beta$ and that each of $n$ infected classes recovers at
rate $\gamma n$, resulting in gamma-distributed infectious
periods.

In this paper we investigate the growth of an epidemic from a single
infection to a full-scale epidemic, without the restrictive assumptions
underlying~\eqref{eqn:Sdot}--\eqref{eqn:Rdot}.  In
\S{}\ref{sec:stochastic}, we show how to model the early
stochastic phase, and give some comparison with deterministic
predictions.  In \S{}\ref{sec:deterministic} we show how to find
deterministic equations governing the epidemic's growth.  We take a
different approach from most previous studies and arrive at a system
similar to the standard equations~\eqref{eqn:Sdot}--\eqref{eqn:Rdot}
rather than a partial differential equation.
If the change in $N/S$ is not large during a typical infectious
period, we can approximate the infectious population as being in
equilibrium given $N/S$ and arrive at a memoryless system that captures
the dynamics well.  
In \S{}\ref{sec:transition} we examine what it means for the
outbreak to be large enough to be deterministic.

\section{Stochastic Phase}
\label{sec:stochastic}

We assume that the disease spreads from individual to individual in
such a way that the ability of individual $u$ to infect a susceptible
individual depends
only on how long $u$ has been infected and whether or not $u$ has
recovered.  We let $P(\tau)$ be the probability $u$ is still infected
$\tau$ units of time after becoming infected.  If $u$ is still infected, the rate $u$
causes new infections is $\beta(\tau)S/N$.  This enforces a possibly
unrealistic assumption that infectiousness is independent of infection
duration.  It would be straightforward to modify the model to
incorporate this effect, but we do not do it here.

We have $P(0)=1$ and --- assuming no-one remains infectious forever ---
$P(\infty)=0$.  We assume $P$ is differentiable.  The
probability of recovering in a short interval $(\tau,\tau+\Delta
\tau)$ is $-P'(\tau)\Delta \tau + \order(\Delta \tau^2)$.  We let
$P_{rec}(\tau)$ be the rate at which recovery happens:
$P_{rec}(\tau)=-P'(\tau) \geq 0$.

\subsection{The equations}
We have full derivations of the equations in appendix~\ref{app:pgf}.
If $p_k(t)$ is the probability that $k$ individuals are infected at
time $t$, then the probability generating function (pgf) $f(x,t) =
\sum_{k=0}^\infty p_k(t)x^k$ provides a useful tool to help calculate $p_k$.
We get
\begin{equation}
f(x,t) = xP(t)g(x,t|t) + \int_0^t g(x,t|\tau) P_{rec}(\tau) \, \mathrm{d}\tau
\label{eqn:f}
\end{equation}
Here $g(x,t|\tau)=\sum q_k(t|\tau)x^k$ is the pgf for the number of
descendants an individual has $t$ units of time after its infection given
that it recovers $\tau\leq t$ units of time after infection.  That is
$q_k(t|\tau)$ is the probability an individual has $k$ infectious descendants $t$
units of time after becoming infected given that it recovers after
$\tau\leq t$ units of time.

We find (for $\tau \leq t$)
\begin{equation}
g(x,t|\tau) = \exp\left(\int_0^\tau [f(x,t-\theta)-1]\beta(\theta)
  \, \mathrm{d}\theta\right)
\label{eqn:g}
\end{equation}
  To find
equations for $p_k(t)$ we take the $k$-th derivative of $f$, divide by
$k!$, and evaluate at $x=0$.  We solve the equations as described in
appendix~\ref{app:earlynumerics}.

\begin{figure}
\includegraphics[width=0.3\columnwidth]{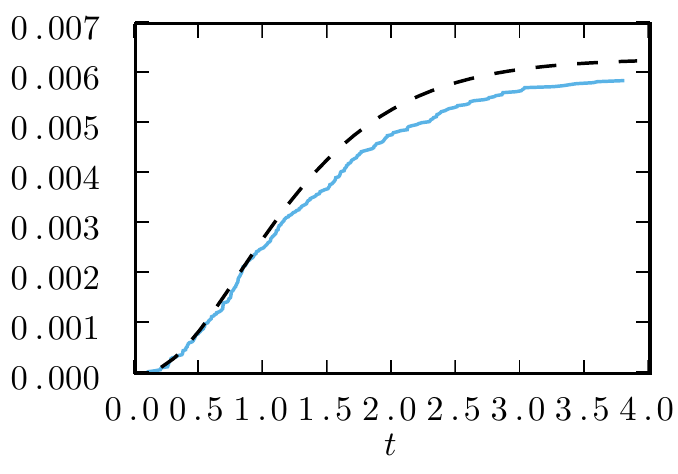}\hfill
\includegraphics[width=0.3\columnwidth]{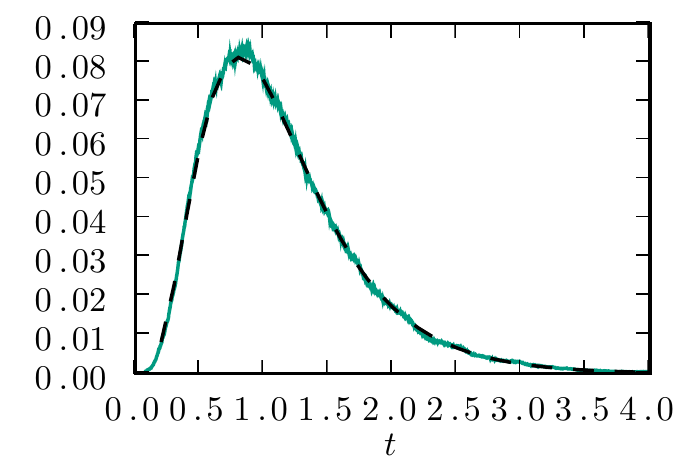}\hfill
\includegraphics[width=0.3\columnwidth]{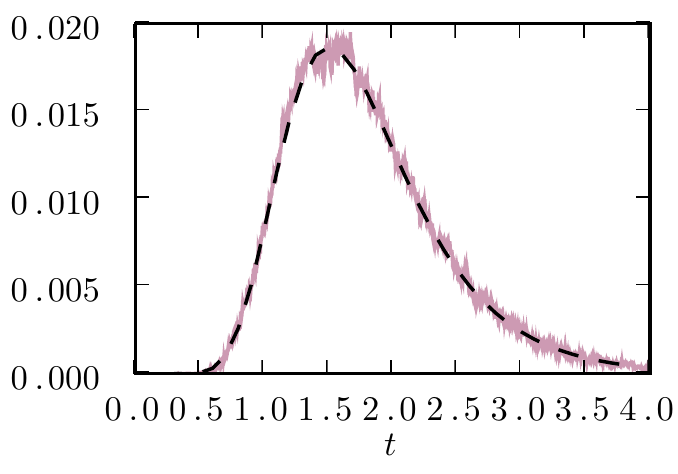}
\caption{The probability of having $0$, $5$, or $20$ people infected
  as functions of time beginning with a single index case: comparison
  of theory (dashed) and $50\,000$ simulations.}
\label{fig:manp}
\end{figure}

We compare the solutions with $50\,000$ simulations in
figure~\ref{fig:manp}.  We take the probability distribution function (pdf) of the infection duration to
be a Weibull distribution, $W(5.8,2.59)$, so $P(\tau) =
e^{-(\tau/5.8)^{2.59}}$.  We take constant $\beta=2$.  Although there
is considerable noise in simulations, we find
close match with analytic results.

\subsection{Asymptotic behavior at large $I$}
If $S(0)$ is large, then $N/S$ may still be approximately
constant even as $I$ becomes much larger than $1$.  We are interested
in the behavior of $I$ as it becomes large, but before $N/S$ has
changed significantly.  If we assume $N/S=1$ remains fixed, then under
weak assumptions it can be shown~\cite{CM1,CM2} that $I(t)$ either
becomes zero at some finite time or it converges to $Ce^{\phi t}$
where $C$ is a random number determined by stochastic effects
and $\phi$ solves
\[
1 = \int_0^\infty e^{-\phi \tau} \beta(\tau)P(\tau) \, \mathrm{d}\tau
\]
This equation is the \emph{Euler-Lotka} (EL) equation, which we derive
in \S{}\ref{sec:deterministic}.  The solution $\phi$ is unique and
known as the \emph{Malthusian parameter}.  The most significant
assumption we require for this convergence is that the infection is
not a ``lattice'' process, that is, possible times of infection are
not discretized and so $I$ can change change
continuously~\footnote{For lattice processes, similar results apply
  with discrete rather than continuous time.}.  This result guarantees
that if the susceptible population is sufficiently large, the outbreak
either dies out or becomes large enough that the growth is
deterministic.

We have shown that equations~\eqref{eqn:f} and~\eqref{eqn:g}
accurately predict the probability of having a given number of
infections as a function of time.  Once the outbreak is sufficiently
large, the impact of stochastic effects is reduced and the infected
population size scales like $Ce^{\phi t}$ for fixed $\phi$.  The
random value of $C$ determines how much time is available to prepare
for the epidemic.

\subsection{Distribution of epidemic onset times}

We turn to a simpler disease process to investigate the impact of the
stochastic phase on how quickly an epidemic ``takes off''.  We
consider a population with constant infectiousness and exponentially
distributed infection durations (corresponding to a constant recovery
rate).  We compare predictions from the stochastic model with
predictions from the deterministic
equations~\eqref{eqn:Sdot}--\eqref{eqn:Rdot} which are exactly valid
precisely for
this infection process.  We take $\beta = 1.5$ and $\gamma = 1$.

\begin{figure}
\begin{center}
\includegraphics[width=0.48\textwidth]{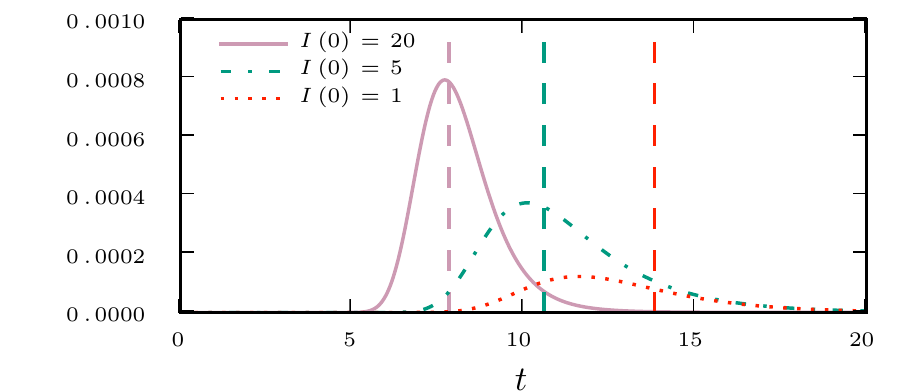}
\end{center}
\caption{A comparison of the deterministically predicted time at which
  $1000$ individuals are infected (vertical dashed lines) with the
  actual probabilities (solid curves) of having $1000$ individuals
  infected at each time given different numbers of initial infections.}
\label{fig:detvsstoch}
\end{figure}

Figure~\ref{fig:detvsstoch} shows that if the initial number of
infections is low, it is relatively likely that the number infected
becomes large before the deterministic equations predict it should.
This has a number of implications for interpreting early stages of an
outbreak.  If we attempt to predict the present size of an outbreak
given a known introduction date using the assumption of deterministic
growth, we are likely to underpredict the current size.  Consequently
if we make preparations to introduce interventions under the
assumption of deterministic growth, we may be using interventions that
are too small and implemented too late.

The mismatch decreases as the initial number of infections increase.
We explain this observation by noting that outbreaks with only a few
infections grow \emph{on average} at the deterministically
predicted rate.  However, those at the lower range of growth often go
extinct, while those at the higher range tend to become epidemics
quickly.  This leads to the important conclusion that if an epidemic
happens, it is likely to happen faster than the deterministic
equations predict.

\section{Deterministic Phase}
\label{sec:deterministic}

In this section we develop the deterministic equations governing
epidemics once stochastic effects are unimportant.  Our exact
equations are equivalent to many previous age-of-infection
models~\cite{breban:aoi,hethcote:aoi,brauer:aoi,brauer:compartmental,li:aoi,castillochavez:aoi,thieme:aoi},
but we avoid the use of PDEs which usually arise.  A related approach
also avoiding PDEs was used by~\cite{brauer:aoi}, but we cast our
equations in a form similar to the standard SIR
equations~\eqref{eqn:Sdot}--\eqref{eqn:Rdot}.  We then introduce an
approximation to these equations. We discuss the transition from the
stochastic phase to the deterministic phase in
\S{}\ref{sec:transition}.

In the stochastic phase analysis, we assumed that infectiousness is
independent of the recovery time (except that after recovery
infectiousness is zero).  We can drop this assumption here and
redefine $\beta(\tau)$ as the average rate of infection $\tau$ units
of time after infection for those individuals still infected (of which
a fraction $S/N$ are successful).  The product $\beta(\tau)P(\tau)$ represents the expected
rate of new infections caused by an individual $u$ infected $\tau$
units of time previously, where the expectation is taken without prior
knowledge of whether $u$ has recovered.  We normalize this by $\Ro =
\int_0^\infty \beta(\tau) P(\tau) \, \mathrm{d}\tau$ to arrive at the
generation interval distribution
$\beta(\tau)P(\tau)/\Ro$~\cite{svensson:generation,wallinga:generation}.

Let $b(t)$ denote the rate of new
infections occurring at time $t$ and $d(t)$ the rate of recoveries.
Let $i(t,\tau)$ denote the number of people who became infected at
time $t-\tau$ and are still infected at time $t$.  Then $i(t,\tau) =
b(t-\tau)P(\tau)$.  We can find $b$ in terms of $i$ by $b(t) =
\int_0^\infty i(t,\tau) (S/N)\beta(\tau) \, \mathrm{d}\tau$ and $d$ in terms of
$b$ by $d(t) = \int_0^\infty b(t-\tau) P_{rec}(\tau) \, \mathrm{d}\tau$.  

If $N/S$ is constant the age-of-infection distribution converges to a
steady-state where $i(t,\tau)/I(t)$ is independent of $t$.  The
population size grows or decays exponentially, so $b(t) = Ce^{t\phi}$
where $\phi$ solves the modified EL equation
\begin{align*}
Ce^{t\phi} &= \int_0^\infty C e^{(t-\tau)\phi} 
  \frac{S}{N} \beta(\tau)P(\tau) \, \mathrm{d}\tau\\
\Rightarrow \frac{N}{S} &= \int_0^\infty e^{-\tau\phi} \beta(\tau) P(\tau) \,
\mathrm{d}\tau
\end{align*}
This has been used at early times~\cite{wallinga:generation} when $N/S
\approx 1$ to relate the exponential growth in time $\phi$ with $\Ro$.

We use the constant $N/S$ solution as the basis for our approach with
changing $N/S$.  We take $b(t)= Ce^{\xi(t)}$
\[
Ce^{\xi(t)} = \int_0^\infty Ce^{\xi(t-\tau)} \frac{S(t)}{N}\beta(\tau)P(\tau) \,
\mathrm{d}\tau
\]
Rearrangement gives
\[
e^{\xi(t)}\frac{N}{S(t)} = \F[\xi,t] \equiv \int_0^\infty  e^{\xi(t-\tau)}
  \beta(\tau) P(\tau)  \, \mathrm{d}\tau
\]

We derive equations for $I$ and $S$ in terms of $\xi$ as follows: The
derivative of $S$ is $-b(t) = -Ce^{\xi(t)}$.  We multiply by $1=
I/\int_0^\infty i(t,\tau) \, \mathrm{d}\tau$, using $i(t,\tau) =
b(t-\tau)P(\tau) = C e^{\xi(t-\tau)} P(\tau)$ to get
\[
\dot{S} = - \frac{Ie^{\xi(t)}}{\G[\xi,t]}=- \frac{\F[\xi,t]}{\G[\xi,t]}\frac{IS}{N}
\]
where $\G[\xi,t] = \int_0^\infty e^{\xi(t-\tau)} P(\tau) \, \mathrm{d}\tau$.
Repeating this for $\dot{I} = b(t)-d(t)$ we get
\[
\dot{I} = \frac{I}{\G[\xi,t]} - \frac{\H[\xi,t]}{\G[\xi,t]} I =
\frac{\F[\xi,t]}{\G[\xi,t]} \frac{IS}{N} - \frac{\H[\xi,t]}{\G[\xi,t]} I
\]
where $\H[\xi,t] = \int_0^\infty e^{\xi(t-\tau)} P_{rec}(\tau) \, \mathrm{d}\tau$.
This can be written in a similar form to the
standard SIR equations, except that the coefficients change in
time and depend on the history of the epidemic
\begin{align}
\dot{S} &= - \hat{\beta}(t)\frac{IS}{N} \label{eqn:fullSdot}\\
\dot{I} &=\hat{\beta}(t) \frac{IS}{N} - \hat{\gamma}(t)
I \label{eqn:fullIdot}\\
\dot{R} &=\hat{\gamma}(t) I \label{eqn:fullRdot}\\
\F[\xi,t] &= \frac{N}{S}e^{\xi(t)} \label{eqn:xi}
\end{align}
where $\hat{\beta}(t)=\F[\xi,t]/\G[\xi,t]$ and $\hat{\gamma}(t) =
\H[\xi,t]/\G[\xi,t]$.  Because of the similarity in notation, we
distinguish $\hat{\beta}(t)$ to be the average rate of causing
infection of all individuals infected at time $t$, while $\beta(\tau)$
is the average rate of causing infection by an individual still
infected $\tau$ units of time after becoming infected.  To initialize
the problem we need $\xi(t)$ for all $t<0$ as well as $S(0)$ and
$I(0)$.  Typically we will assume that $\xi(t)=-\infty$ for $t<0$ so
that $e^{\xi(t)}=0$.  As we solve forward, new values of $\xi$ are
calculated based on the change in $S$.  The history of $\xi(t-\tau)$
for $\tau>0$ encodes all information needed about the age-of-infection
distribution at $t$.  A less intuitive, but simpler
formulation of these equations appears in appendix~\ref{app:equiv}.

\subsection{Approximating the solution}
Storing the history of an outbreak introduces some mild analytical and
computational difficulties.  It is convenient to work with a system that
depends only on its current state.  
If $N/S$ varies slowly relative to how quickly $\xi$ changes, we can
assume that the system responds adiabatically to changes in $N/S$ and
so the age-of-infection distribution is at equilibrium with the
current value of $N/S$.  This assumption will allow us to create
equations analagous to~\eqref{eqn:Sdot}--\eqref{eqn:Rdot} with
changing coefficients, which may be solved by standard ODE methods.
This approach will break down if $N/S$ changes significantly during a
typical infectious period.  Fortunately, we can use the results of the
approximation to identify when the approximation fails.


We replace $\xi(t-\tau)$ by $\xi(t) - \int_0^\tau
\phi(t-\theta) \, \mathrm{d}\theta$ where $\phi(t) = \xi'(t)$ and approximate $\F/e^\xi$,
$\G/e^\xi$, and $\H/e^\xi$ by $\f(\phi)$, $\g(\phi)$, and $\h(\phi)$
respectively assuming that $\phi(t-\tau) \approx \phi(t)$ for the
range of $\tau$ which make a significant contribution to the integral.
\begin{align*}
\f(\phi) &= \int_0^\infty e^{-\tau \phi(t)} \beta(\tau)
P(\tau) \, \mathrm{d}\tau\\
\g(\phi)&= \int_0^\infty e^{-\tau \phi(t)} 
P(\tau) \, \mathrm{d}\tau\\
\h(\phi) &= \int_0^\infty e^{-\tau \phi(t)}  P_{rec}(\tau) \, \mathrm{d}\tau
\end{align*}
Note that each of these is a Laplace transform.  The resulting
approximating equations are
\begin{align}
\dot{S} &= - \hat{\beta}_0(t)\frac{IS}{N} \label{eqn:approxSdot}\\
\dot{I} &=\hat{\beta}_0(t) \frac{IS}{N} -
\hat{\gamma}_0(t) I\label{eqn:approxIdot}\\
\dot{R} &=\hat{\gamma}_0(t) I\label{eqn:approxRdot}\\
\f(\phi) &= \frac{N}{S(t)} \label{eqn:phi}
\end{align}
where $\hat{\beta}_0(t)=\f(\phi)/\g(\phi)$ and $\hat{\gamma}_0(t) =
\h(\phi)/\g(\phi)$.

Computationally this system of equations is only mildly more difficult
than the standard SIR equations.  We can either find the functional
forms of the Laplace transforms, or simply calculate them for various
$\phi$ in advance.  Once that is done, then at each time step, we need
only look at $N/S$, identify $\phi$ such that $\f(\phi)=N/S$, and
then find $G$ and $H$.  Then the integration proceeds as in the
standard SIR equations.

The approximation is valid as long as the amount of change of $N/S$
during a typical infectious period is small, and is therefore valid
well into the nonlinear regime after the exponential growth phase has
ended.

\subsection{Examples}

\subsubsection{The usual suspects}
If we make the usual assumptions of constant infectiousness and
exponentially distributed recovery time [$\beta$ constant and
$P_{rec}(\tau)=\gamma e^{-\gamma\tau}$] the system is memoryless.  The
function $\xi$ encodes the age-of-infection distribution, which is
irrelevant in a memoryless system.  Thus the equations for $I$ and $S$
should not depend on $\xi$.  We find $\F[\xi,t] = \beta \G[\xi,t]$,
and so $\dot{S} = -\beta IS/N$.  We similarly find
$\H[\xi,t]/\G[\xi,t] = \gamma$ and so $\dot{I} = \beta IS/N - \gamma
I$.  So in this special case the exact age-of-infection
model~\eqref{eqn:fullSdot}--\eqref{eqn:xi} reduces to the standard SIR
equations~\eqref{eqn:Sdot}--\eqref{eqn:Rdot}.  This holds even for our
approximate system~\eqref{eqn:approxSdot}--\eqref{eqn:phi}.

\subsubsection{A piecewise continuous example}
\label{sec:pc}
\begin{figure}
\includegraphics[width=0.3\textwidth]{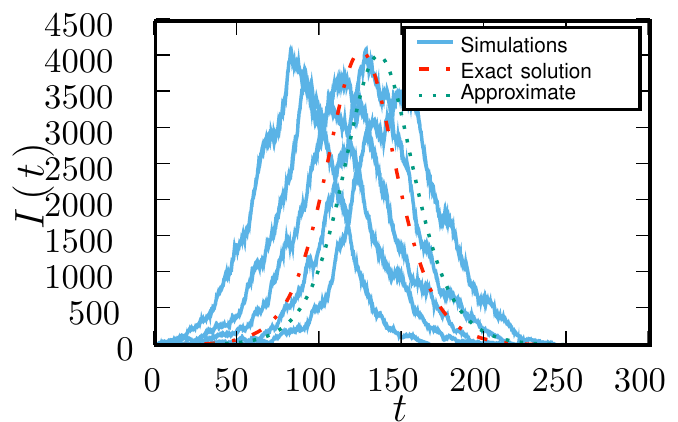}
\hfill
\includegraphics[width=0.3\textwidth]{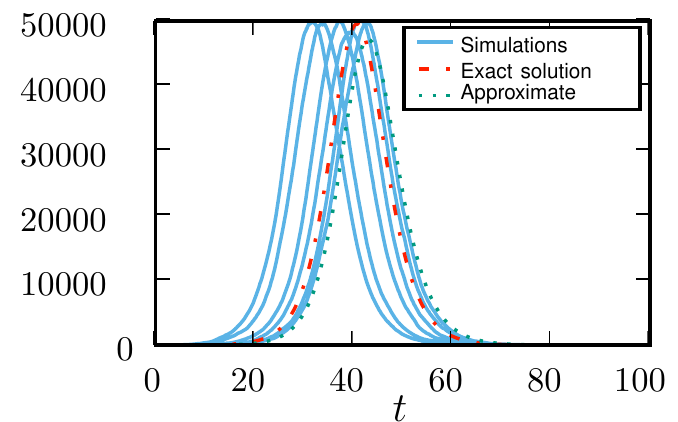}
\hfill
\includegraphics[width = 0.3\textwidth]{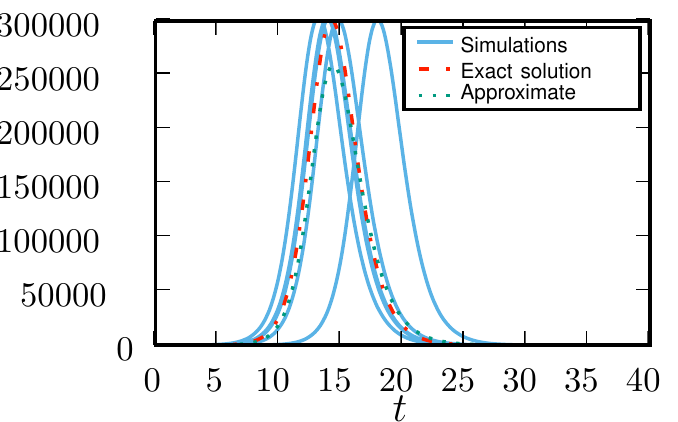}
\caption{Comparison of simulations with exact age-of-infection model,
  approximation, and two parametrizations of the SIR equations.  The
  temporal shift of the exact and approximate solutions is a result of
  difference in initial condition.  The exact solution takes the
  initial condition that $\xi(t)=0$ for $t<0$ while the approximate
  solution assumes that $\xi(t)=t\phi(0)$ for $t<0$.}
\label{fig:nasty}
\end{figure}
We take 
\begin{align}
\beta(\tau) &= \begin{cases} c & 0 \leq \tau \leq 1 \text{ or } 2 \leq
  \tau \leq 3 \\
 0 & \text{otherwise}
\end{cases} \label{eqn:PC_beta}
\\
P_{rec}(\tau) &= \begin{cases} 1/2 & 1 \leq \tau \leq 3\\ 
                0 & \text{otherwise} 
\end{cases} \label{eqn:PC_Prec}
\end{align}
So people are initially infectious, then stop being infectious at
$\tau=1$ and begin to recover.  At $\tau=2$, they continue recovering,
but become infectious once more.  By $\tau=3$ all individuals have
recovered.  Such a system could model a disease in which individuals
are infectious before and possibly after having symptoms, but
self-isolate during the symptomatic phase.  The generation interval
distribution is given by
\begin{equation}
\frac{\beta(\tau)P(\tau)}{\Ro} = \begin{cases} 4/5 & 0 \leq \tau \leq 1\\  
 \frac{2(3-\tau)}{5} & 2 \leq \tau \leq 3\\
0 & \text{otherwise}
\end{cases}
\label{eqn:PC_GenInt}
\end{equation}

In figure~\ref{fig:nasty} we find that the exact
model~\eqref{eqn:fullSdot}--\eqref{eqn:xi} fits the simulations well
(with the discrepancy due to stochastic shifts in time).  The
difference in timing between the exact and
approximate solution~\eqref{eqn:approxSdot}--\eqref{eqn:phi} is due to
differences in initial conditions: the exact calculation assumes a
single infection beginning at $t=0$ while the approximate solution
assumes that the epidemic begins with the equilibrium age-distribution
already reached by $t=0$.  The approximate model is a good fit for the
behavior at early times and remains a good approximation until the
change in $N/S$ becomes significant over the duration of an infection.
The approximation performs best in those situations where the number of
infections remains smaller.

\begin{figure}
\includegraphics[width=0.3\textwidth]{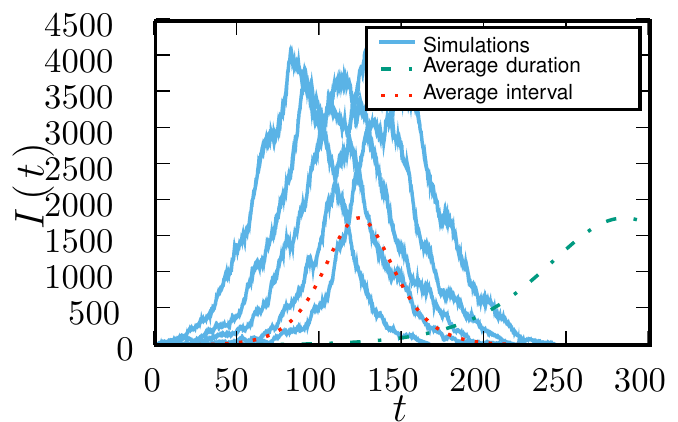}
\hfill
\includegraphics[width=0.3\textwidth]{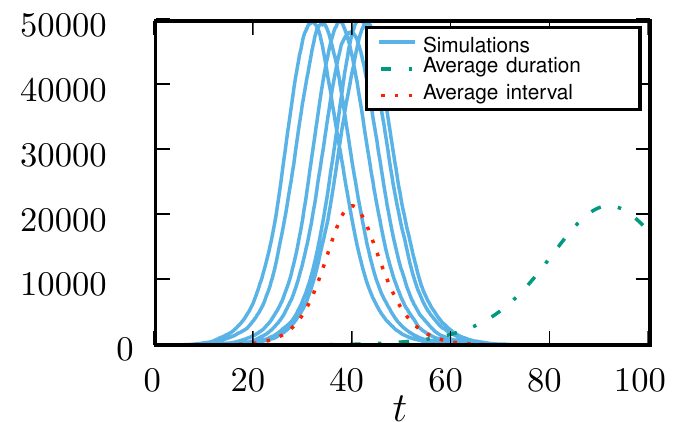}
\hfill
\includegraphics[width = 0.3\textwidth]{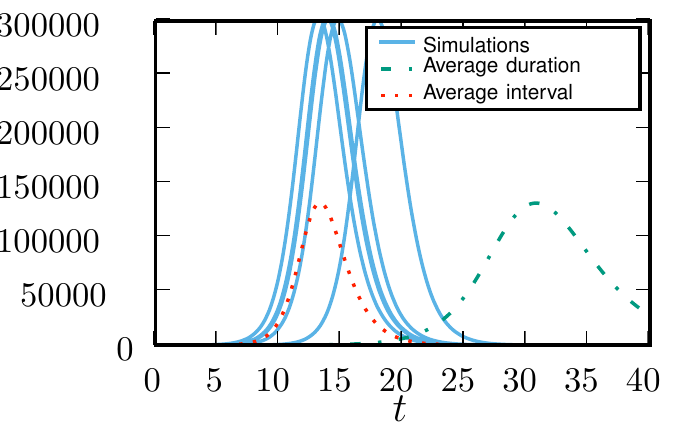}
\caption{The standard SIR equations cannot closely capture the
  dynamics of the disease spread, regardless of whether we preserve
  the average duration of infection or the average generation interval.}
\label{fig:nasty_naive}
\end{figure}
If we attempt to approximate the epidemic course using the standard
SIR model~\eqref{eqn:Sdot}--\eqref{eqn:Rdot}, then we have two free
parameters $\beta$ and $\gamma$.  We can identify (at least) three
constraints: $\Ro$, average duration of infection, and average
generation interval.  We can only match two of these at a time, which
we show in figure~\ref{fig:nasty_naive}.  If we choose to match $\Ro$
and average duration of infection then the total number of infected
person-days is correct, but the timing is far off.  If we choose
to match $\Ro$ and average generation interval, then the timing is
much closer, but the peak patient load is significantly
underestimated.




\subsubsection{Gamma-distributed recovery times}

\begin{figure}
\includegraphics[width=0.3\textwidth]{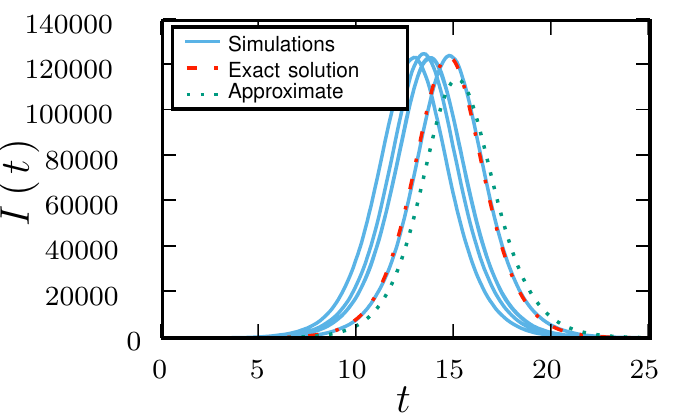} \hfill 
\includegraphics[width=0.3\textwidth]{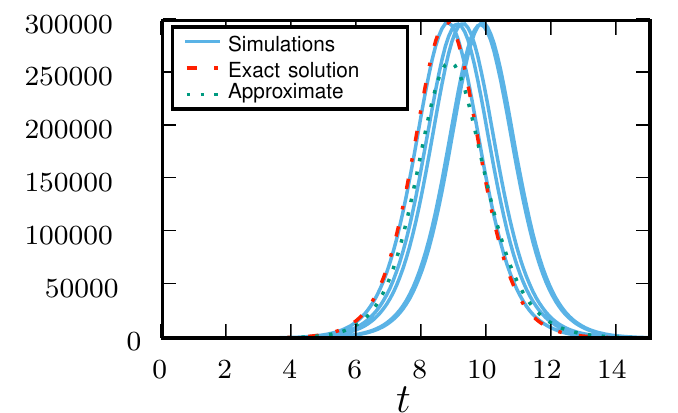} \hfill 
\includegraphics[width=0.3\textwidth]{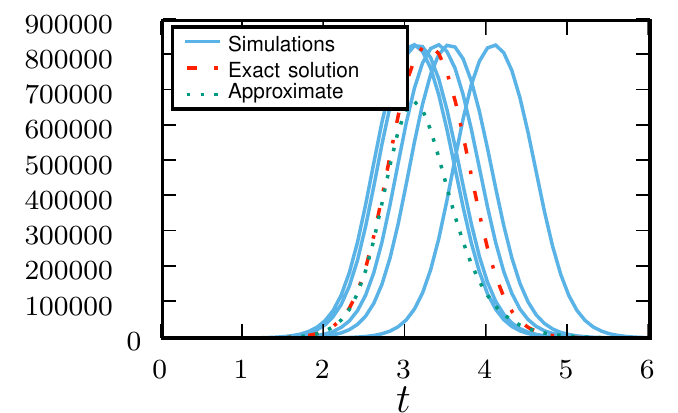} \hfill 
\caption{For gamma distributed recovery time with constant
  infectiousness, the exact system differs from simulations only in
  time shifts.  The approximation closely matches the initial growth
  phase, but begins deviating close to the peak.}
\label{fig:gamma}
\end{figure}
Recently~\cite{wearing} investigated some of the role the distribution
of infection duration has on the dynamics of an epidemic.  They
considered a gamma-distributed infectious period with constant
infectiousness.  The model they studied corresponds to a chain of
$100$ exponentially distributed infectious classes, each with
infectiousness $\beta$ and expected duration $1/100$.  They showed that
the standard SIR equations~\eqref{eqn:Sdot}--\eqref{eqn:Rdot} provide
a poor approximation.

For this system, $P_{rec}(\tau) = \tau^{n-1}\exp(-n\tau)n^n/(n-1)!$
where $n=100$.  The Laplace transform of this is $(1+\phi/n)^{-n}$.
From this we can derive the transforms of $P$ and $\beta P$, which
allows us to define the coefficients for our approximation.
Figure~\ref{fig:gamma} shows that the approximation closely follows
the early growth even after the exponential phase ends.  It finally
deviates close to the epidemic peak, but it gives a reasonable
estimate of the timing and maximum load of the epidemic.
\section{Transition Phase}
\label{sec:transition}
We have shown that stochastic effects play an important role on
whether an epidemic occurs and the timing of an epidemic if it does
occur.  We have also seen that once the epidemic is sufficiently
large, it follows the deterministic predictions.  We borrow an
approach from~\cite{gillespie:chemical} to identify when the
transition from the stochastic phase to the deterministic phase
occurs.
For simplicity in our analysis, we will assume that the process is not
highly peaked.  This allows us to assume that $i(t,\tau)/I(t)$ is close to
its equilibrium state.  

In order to treat the dynamics as deterministic over a time interval
$\Delta t$, we must satisfy two competing conditions.  First, we need
the time interval to be large enough that the number of infections and
recoveries that happen in that interval is well-approximated by the
expected number.  That is, we need the expected error to be small
compared to the expected value, and so the \emph{coefficient of
  variation} (the square root of the variance divided by the
expectation) is small.  Assuming that the rates remain constant, the
infection and recovery processes are both Poisson, and so their
difference is a Skellam distribution, which has variance $I\Delta
t(\hat\beta + \hat\gamma)$~\cite{skellam,johnson:univariate}.
Consequently the condition we need is that $\sqrt{I\Delta t(\hat\beta
  + \hat\gamma)}/I\Delta t |\hat\beta-\hat\gamma| \ll 1$.  So
\begin{equation}
\Delta t \gg
\frac{\hat\beta+\hat\gamma}{I(\hat\beta-\hat\gamma)^2}
\label{eqn:transcond1}
\end{equation}
Second, we need the time interval to be small enough that the rate at
which the infectious population size changes is not affected by changes in
the infectious population.  That is we need $\Delta I
\approx (\beta-\gamma) I \Delta t \ll I$.  So
\begin{equation}
\Delta t \ll \frac{1}{|\hat\beta-\hat\gamma|}
\label{eqn:transcond2}
\end{equation}
For small values of $I$, conditions~\eqref{eqn:transcond1} and~\eqref{eqn:transcond2} cannot be satisfied
simultaneously.  

Combining these conditions we need that
\[
I \ggg \frac{\hat{\beta} + \hat{\gamma}}{|\hat{\beta}-\hat{\gamma}|}
\]
More strictly, we actually require that $\sqrt{I} \ggg
(\hat{\beta}+\hat{\gamma})/|\hat{\beta}-\hat{\gamma}|$.

The analysis we have done does not apply close to the peak of the
epidemic (where $\hat{\beta} = \hat{\gamma}$).  Here we can replace
condition~\eqref{eqn:transcond1} with the requirement that the error
in the number of new infections is small compared to the number of
new infections and similarly for the number of recoveries.  In general we
need condition~\eqref{eqn:transcond2} combined with either this pair
of conditions or condition~\eqref{eqn:transcond1} to guarantee that
the deterministic equations apply.  For practical purposes, once the
deterministic equations hold, we expect them to hold through the peak
until $I$ decays at which point we can use~\eqref{eqn:transcond1}
again.

If the generation interval distribution were highly peaked around some
typical time, then we could still argue that the system is
deterministic, but we would have to explicitly set the history of
$\xi$ rather than assuming it is takes the equilibrium form.  By
assuming the equilibrium distribution we can treat infections as
occurring at a slowly changing rate.



\section{Discussion}
A typical disease outbreak begins small and whether it grows or
becomes extinct is strongly influenced by stochastic effects.  If it
grows, it generally does so faster than predicted deterministically
because those outbreaks which are most likely to not die out
stochastically are those which initially grow faster than average.
Consequently if we observe an epidemic, it is likely to have grown to
an epidemic faster than deterministic equations predict.

Once an outbreak becomes large, it transitions to a deterministic
phase.  We can estimate the size an outbreak must reach to be
deterministic by identifying a time interval which is large enough
that many events happen in the interval (and so the error of a
deterministic prediction is small compared to the prediction), while
at the same time the interval is small enough that the size of $I$ and
$S$ do not change significantly.  Such a time interval can only exist
if $I$ is sufficiently large.

Once an outbreak is deterministic, we can use the deterministic
equations to accurately model the spread once a correcting time shift
is applied.  These equations are somewhat difficult because they
require saving the history of an epidemic, and so it may be more
convenient to use approximate models.  We have introduced an
approximate model based on standard compartmental models.  We
assume that the system responds adiabatically to changes in the
susceptible fraction.  It uses a single infectious class, but has
coefficients that change in time.  It provides a good estimate of the
early behavior, but may deviate close to the peak.  We can estimate
when it deviates by looking at how quickly the susceptible fraction
changes during a typical infectious period.

We have assumed throughout that the infectious population can be
modeled in continuous time.  If the generation interval is discrete,
then these assumptions fail, but similar approaches work in
discrete time.  A more complicated situation arises when the
generation interval distribution is close to discrete: If the
distribution is tightly peaked about a mean which is sufficiently far
from zero, then it may take many generations for the infectious
population to reach equilibrium.  The population may become
deterministic before the population reaches equilibrium, in which case
our exact equations will provide a good model (assuming appropriate
initial conditions) while our first approximation may fail badly.  Our
second approximation may require a long chain of infectious classes in
order to reproduce the correct dynamics.

The models we have developed are straightforward to adapt to SIR with
birth or death, SIS, or SIRS.  In fact, such situations will be more
amenable to our first approximating method because the rate of change
of $N/S$ is reduced.

\section*{Acknowledgments}

BP would like to acknowledge the support of Canadian Institutes of
Health Research (CIHR) (grants nos. PTL-93146, PAP-93425 and
MOP-81273) and the Michael Smith Foundation for Health Research
(MSFHR) (Senior Scholar Funds).  JCM, BD, and RM were supported by
these grants.

JCM was additionally supported by the RAPIDD program of the Science \&
Technology Directorate, Department of Homeland Security and the
Fogarty International Center, National Institutes of Health.

\appendix
\section{Probability Generating Functions}
\label{app:pgf}
A probability generating function (pgf) is a function $f(x)$ which
encodes a probability distribution of non-negative integers~\cite{gf}.
Given that the probability of $k$ is $p_k$ we define the function
\[
f(x) = p_0 + p_1x^1 + p_2x^2 + \cdots
\]
Probability generating functions have a number of useful properties.
The product of two pgfs is itself a pgf for the sum of two numbers
chosen from each distribution.  From this fact, it can be shown that
for two pgfs $f$ and $g$ encoding the distributions $P_g$ and $P_f$
respectively, the function $f(g(x))$ is the pgf for the distribution
found by choosing a random number $s$ from $P_f$, and then taking the
sum of $s$ random numbers from $P_g$.

This property of function composition is useful in our context to
deal with taking a random number of infecteds (corresponding to
$P_f$), and each of them infects a random number of susceptibles (from
a distribution $P_g$).  The resulting number of new infections is
given by the composition of the corresponding pgfs.

\subsection{derivation of equations}
We assume that the population is sufficiently large relative to the
number of infections, that no infections are prevented by depletion of
susceptibles.  We focus our attention on a single infected individual
$u$ and its descendants.  We can assume that $t=0$ when $u$ becomes
infected.  Let $f(x,t)$ be the time-dependent pgf for the number of
individuals (descended from $u$, including $u$) who are infected at
$t$.  That is $f(x,t) = \sum_{n=0}^\infty p_n(t)x^n$ where $p_n(t)$ is
the probability that $n$ individuals are infected at time $t$.

Let $g(x,t|\tau)$ be the pgf for the number of infectious descendants
$u$ has $t$ units of time after becoming infected given that its
infection lasts $\tau$ units of time.  Note that if $\tau>t$, then
$g(x,t|\tau) = g(x,t|t)$.  Then the number of current infections is
given by a weighted average of the number of descendants (plus $1$ if
$u$ is still infectious).  Encoding this as a statement for pgfs gives
\[
f(x,t) = xP(t)g(x,t|t) + \int_0^t g(x,t|\tau) P_{rec}(\tau) \, \mathrm{d}\tau
\]
The number of infections resulting from an individual $v$ infected at
time $\theta$ has pgf $f(x,t-\theta)$.  This allows us to express $g$
in terms of $f$.

To find $g$, we consider an individual who recovers at time $\tau$ and divide the
duration of infectiousness into small $\Delta \theta$ sized blocks.
The pgf for the number of infections at
time $t$ due to an infection that occurs in the interval $[\theta,
\theta + \Delta \theta)$ is  $f(x,t-\theta) + \order(\Delta \theta)$.  The infection
occurs with probability $\beta(\theta) \Delta \theta + \order(\Delta \theta^2)$.
The probability that infection does not occur during that time period
is $1-\beta(\theta)\Delta \theta + \order(\Delta \theta^2)$.  Consequently the
pgf for the number of infections at time
$t$ resulting from infections in the time interval of interest is:
\[
1 + [f(x,t-\theta)-1] \beta(\theta) \Delta \theta + \order(\Delta \theta^2)
\]
The pgf for the number of
infections occurring in any of the time intervals is the product of the
individual generating functions.  Consequently, taking $\Delta \theta
\to 0$, the pgf for the number of
descendants an individual has at time $t$ given that it recovers at $\tau
\leq t$ is
\begin{align*}
g(x,t|\tau) &=\lim_{\Delta \theta \to 0} \prod_{i=0}^{\tau/\Delta \theta} 1 + [f(x,t-i\Delta \theta) -1]\beta(i\Delta \theta) \Delta \theta + \order(\Delta \theta^2)\\
&= \lim_{\Delta \theta \to 0} \exp(\sum_{i=0}^{\tau/\Delta\theta} \ln (1+[f(x,t-i\Delta \theta) - 1]\beta(i\Delta \theta)\Delta \theta + \order(\Delta\theta^2)\\
&=\lim_{\Delta \theta\to 0} \exp\left( \sum_{i=0}^{\tau/\Delta \theta} [f(x,t-i\Delta \theta)-1] \beta(i\Delta \theta) \Delta \theta + \order(\Delta \theta^2)\right)\\
&=\exp\left(\int_0^\tau [f(x,t-\theta)-1]\beta(\theta) \, \mathrm{d}\theta\right)
\end{align*}
If the individual recovers at time $\tau > t$, then the pgf for the
number of descendants at time $t$ including itself satisfies $g(x,t|\tau) =
xg(x,t|t)$.

This expression for $g$ can be derived alternately by considering a
large population size $N$ and noting that if the expected number of
infections caused by $v$ is $r = \int_0^\tau \beta(\theta)
\, \mathrm{d}\theta$, then the probability of infecting each individual is $p =
\int_0^\tau \beta(\theta)/N$  The probability of infecting $n$ people
is then $\binom{N}{n} p^n(1-p)^{N-n}$.  From this we can derive the
pgf for the number of infections caused directly from $v$, and then
using function composition will arrive at the same expression.

\section{Notes on the numerics for the stochastic problem}
\label{app:earlynumerics}

We take $f(x,t) = \sum p_k(t)x^k$ and $g(x,t|\tau) = \sum
q_k(t|\tau)x^k$ where $p_k$ gives the probability of having $k$ people
infected at time $t$, while $q_k$ gives the probability of having $k$
descendants given that recovery occurs at time $\tau$.  If we take $k$
derivatives of these equations, divide by $k!$ and evaluate at $x=0$,
we get the probability of $k$ infections.  The resulting system of
equations is straightforward to solve numerically.  As our initial
condition at $t=0$ we generally set all derivatives of $f$ to be $0$
except the first derivative, which is $1$, though other options are
possible.

If we make a simplifying assumption that $\beta$ is constant, we can
find an expression for $g$ which reduces the dimenionality of the
problem.  We have
\begin{align*}
\int_0^\tau [f(x,t-\theta)-1]\beta \, \mathrm{d}\theta &= \beta \left[-\tau + \int_0^t
f(x,t-\theta) \, \mathrm{d}\theta - \int_{\tau}^t f(x,t-\theta) \, \mathrm{d}\theta \right] \\ 
&=-\beta \tau + \beta[\int_0^t
f(x,\theta) \, \mathrm{d}\theta - \int_0^{t-\tau} f(x,\theta) \, \mathrm{d}\theta]
\end{align*}
We define the auxiliary function $\zeta(x,s) = \int_0^s f(x,\theta) \,
\mathrm{d}\theta$.  Then
\[
g(x,t|\tau) = \exp \beta[\zeta(x,t)-\zeta(x,t-\tau) - \tau]
\]
Our equation for $f$ remains
\[
f(x,t) = xP(t)g(x,t|t) + \int_0^t g(x,t|\tau) P_{rec}(\tau) \, \mathrm{d}\tau
\]
This allows us to simplify the calculations by storing $\zeta$ at each
value of $s$ rather than needing to integrate $f$ at each time step.

In practice we want to find arbitrary derivatives of $f$ evaluated at
$x=0$.  To find this numerically, we differentiate these equations
with respect to $x$ to arrive at equations coupling derivatives of
$f(x,t)$ with derivatives of $\zeta$ at $x=0$.  Let us assume we know
$\zeta(0,s)$ and its derivatives for $s=0,dt,2dt,...,t$ and $f(0,t)$
and its derivatives.  To find $\zeta(0,t+dt)$ and $f(0,t+dt)$, it is
straightforward to use an implicit numerical method.

\section{An equivalent formulation}
\label{app:equiv}
Although equations~\eqref{eqn:fullSdot}--\eqref{eqn:xi} are intuitively appealing because of their
similarity to the standard SIR equations, we can reduce them to a
simpler form.  We first replace $e^{\xi(t)}$ with $\psi(t)$.  Note that $\dot{S} = -b(t) = -C\psi(t)$.
Further $\G = I(t)/C$, so from the initial condition at $t=0$,
we can calculate $C$, and have no further need for $g$.  Thus we
arrive at
\begin{align}
\dot{S} &= -C\psi(t)\\
\dot{I} &= C\psi(t) - C\int_0^\infty \psi(t-\tau) P_{rec}(\tau) \, \mathrm{d}\tau\\
\dot{R} &= C\int_0^\infty \psi(t-\tau) P_{rec}(\tau) \, \mathrm{d}\tau\\
\F[\psi,t] &= \frac{N}{S} \psi(t)
\end{align}
If we take as the initial condition that all infections at time $t=0$
begin their infection period at $t=0$, then $\psi(t-\tau)=0$ for
$\tau>t$ and we can assume that the integrals have their upper limit
at $\tau=t$.  If we take some other initial condition, we may have to
include the entire range of $\tau$.  Although these equations are
simpler to solve, they lose some of their intuitive appeal because it
is more difficult to identify the meaning of each term.

\end{document}